\documentstyle[11pt,amssymb,amsfonts,amsthm,amssymb,amscd,amstext,epsfig,amsmath]{article}

\newcommand{\startappendix}{
\setcounter{section}{0}
\renewcommand{\thesection}{\Alph{section}}}
\newcommand{\Appendix}[1]{
\refstepcounter{section}
\begin{flushleft}
{\large\bf Appendix \thesection: #1}
\end{flushleft}}

\def\be{\begin{equation}}
\def\ee{\end{equation}}
\def\ba{\begin{array}}
\def\ea{\end{array}}
\def\del{\partial}

\def\dalemb#1#2{{\vbox{\hrule height .#2pt
        \hbox{\vrule width.#2pt height#1pt \kern#1pt
                \vrule width.#2pt}
        \hrule height.#2pt}}}

\newcommand{\bea}{\begin{eqnarray}}
\newcommand{\eea}{\end{eqnarray}}

\newcommand{\Tr}{{\rm Tr} }

\numberwithin{equation}{section}

\begin{document}
\begin{flushright}
\hfill{DAMTP-2005-103} \\
{hep-th/0511164}
\end{flushright}

\begin{center}
\vspace{1cm}
{ \LARGE {\bf Precision Test of AdS/CFT in Lunin-Maldacena Background}}

\vspace{1.5cm}

Heng-Yu Chen${}^1$ and S. Prem Kumar${}^{2}$
\vspace{0.8cm}

{\it ${}^1$ DAMTP, Centre for Mathematical Sciences,
Cambridge University\\ Wilberforce Road, Cambridge CB3 OWA, UK}

\vspace{0.3cm}

{\it ${}^2$ Department of Physics, University of Wales Swansea\\
Swansea, SA2 8PP, UK}

\vspace{0.3cm}

h.y.chen@damtp.cam.ac.uk \hspace{1cm} s.p.kumar@swan.ac.uk

\vspace{2cm}

\end{center}

\begin{abstract}

We obtain the solutions and explicitly calculate the energy for a
class of two-spin semiclassical string states in the Lunin-Maldacena
background. These configurations are the $\beta$-deformed
versions of the folded string solutions in $AdS_5\times S^5$ background.
They correspond to certain single trace operators in the ${\cal N}=1$
superconformal $\beta$ deformation of ${\cal N}=4$ Yang-Mills.
We calculate the one loop anomalous
dimension for the dual single trace operators from the associated
twisted spin chain with a general two-cut distribution of Bethe roots. Our
results show a striking match between the two calculations. 
We
demonstrate the natural identification of parameters on the two sides of the
analysis, and explain the significance of the Virasoro constraint
associated with the winding motion of semiclassical strings 
from the perspective of the spin chain solution.

\end{abstract}

\pagebreak
\setcounter{page}{1}

\section{Introduction}

The AdS/CFT correspondence
\cite{Maldacena1,Witten1,Gubser1,Aharony} in its original form states the
equivalence between Type IIB superstring theory propagating on 
$AdS_{5}\times S^{5}$ and ${\mathcal{N}}=4$, $SU(N)$ supersymmetric
Yang-Mills (SYM) theory in the large $N$ limit, defined on the boundary of
$AdS_{5}$. Extending this duality to situations with lower
supersymmetry (SUSY) is most naturally and directly achieved by
considering deformations of ${\cal N}=4$ theory which translate to, via
the precise dictionary of the original correspondence, specific
deformations of the $AdS_{5}\times S^{5}$ background. A particularly
interesting class of such deformations are the exactly marginal
ones preserving ${\cal N}=1$ SUSY, elucidated by Leigh and Strassler
in \cite{Leigh:1995ep} and mentioned by earlier authours \cite{early}. One of these deformations, the
so-called $\beta$ deformation, and its large $N$ gravity dual
constructed by Lunin and Maldacena \cite{Lunin:2005jy} will be the
focus of this paper. 

In particular, we will look at certain semiclassical
string solutions in the Lunin-Maldacena background, which are the
$\beta$-deformed versions of the two-spin folded string solutions of
\cite{Frolov:2003xy} in $AdS_5\times S^5$. We will show that the
energy of such configurations precisely matches with the one-loop
corrected scaling dimension of the corresponding gauge theory
operator, computed using the twisted XXX $SU(2)$ spin-chain.
 
The $\beta$ deformation can be understood rather simply as a
modification of the superpotential of the ${\cal N}=4$ theory via
\be
W=\Tr[\Phi_1,[\Phi_2,\Phi_3]]\rightarrow\kappa
\Tr[e^{i\pi\beta}\Phi_1\Phi_2\Phi_3-e^{-i\pi\beta}\Phi_1\Phi_3\Phi_2].
\ee
This deformation yields an ${\cal N}=1$ superconformal field theory
(SCFT) and preserves a $U(1)\times U(1)\times U (1)_R$ global
symmetry subgroup of the $SO(6)_R$ symmetry of the ${\cal N}=4$
theory. These global symmetries translate to $U(1)^3$ isometries in
the dual geometry of \cite{Lunin:2005jy}, which was central to their
construction of the supergravity (SUGRA) solution. We obtain classical
string solutions which are pointlike in the $AdS_5$ part of the geometry,
and which carry
large angular momenta $J_1$ and $J_2$ along two of the $U(1)$ isometry
directions. They are the $\beta$-deformed analogs (with $\beta\ll1$)
of the folded string 
solutions of \cite{Frolov:2003xy}, and while being ``folded'' solutions
they also wind around one of the $U(1)$ isometry directions.  
These solutions are dual to gauge theory operators of the
type $\Tr[\Phi_1^{J_1}\Phi_2^{J_2}]+{\rm permutations}$, with
$J_{1,2}\gg1$. Such operators are non-BPS operators and thus their
scaling dimensions can and do get quantum corrections. Importantly,
the range of $\beta$ 
which we specify below and consequently the class of states that we
consider, are actually distinct from those studied recently in
\cite{Frolov:2005ty,Frolov:2005dj}.

Thus the operators we are looking at are far from BMN operators
\cite{Berenstein:2002jq} as they have large numbers of
``impurities''.
We make use of the remarkable observation made in \cite{Minahan:2002ve}
which revealed that 
the field theory problem of calculating the one loop anomalous
dimensions 
of large operators such as the ones we are considering, can
be reduced to the problem of diagonalising certain spin chain
Hamiltonians using the well-known Bethe Ansatz. 
In the context of the ${\cal N}=4$ theory, striking agreements between 
semiclassical spinning string energies in $AdS_5\times S^5$ and
anomalous dimensions of operators have been demonstrated in 
\cite{Frolov:2003xy}, \cite{Frolov:2003qc}-\cite{Beisert:2004ry}. 
The beautiful
underlying integrable structures which underpin the entire ``spin
chain/spinning string'' correspondence have been revealed in
a series of papers 
\cite{Bena:2003wd}-\cite{Beisert:2004ag}. 
Moreover, the idea of integrabilities has also been applied in the study of certain defect theory \cite{Okamura:2005cj}.
Finally, there are also interesting recent attempts to
semiclassically quantize the entire string spectrum derived from
certain subsectors of the string sigma model \cite{Dorey:2005zz}.  

Extending this remarkable spin-chain/spinning string
correspondence to the less supersymmetric setup of ${\cal N}=1$,
$\beta$ deformed theories could be of
phenomenological interest, but more immediately, it might also tell us
whether the above
integrable structures are merely characteristics of the maximally
supersymmetric theory or a more general feature of
gauge/string dualities, similar to the integrablity in the
correspondence between two dimensional string theory and
matrix quantum mechanics. 

The key point that makes the $\beta$ deformed theories special in this
context is that they are obtained by a continuous, {\em exactly}
marginal deformation 
of the maximally supersymmetric theory. The resultant field theory is
superconformal for all values of the gauge coupling parameter and
importantly,
exists at weak gauge couplings which allows perturbative calculations. 
The construction of the dual SUGRA background by \cite{Lunin:2005jy}
  has made it possible to further explore properties of the string
  dual of the $\beta$ deformation
  \cite{Frolov:2005ty,Frolov:2005dj}. (The techniques of 
 \cite{Lunin:2005jy} have also recently been applied to generate
the deformations of ${\mathcal{N}}=1,2$
theories and there are also many interesting related results \cite{Gursoy:2005cn}).

The integrability of the dilatation operator in the $\beta$ deformed
theory was first discussed in 
\cite{Berenstein:2004ys}, where the one-loop anomalous dimension
matrix for real $\mathcal{\beta}$ was found to be the ``twisted'' XXX
spin chain Hamiltonian. We will use the $SU(2)$ version of this
Hamiltonian to compute the anomalous dimesions of the operators 
$\Tr[\Phi_1^{J_1}\Phi_2^{J_2}]$ for $\beta\ll1$. Let us now be more
specific about the range of $\beta$ and the class of states we are
studying. 

One of our primary motivations is to understand the states in the
theory with small $\beta\ll1$. This question is particularly
interesting for the theories with  
$\beta=1/n$, where $n$ is an integer and $n>>1$. As we discuss in
detail in Section 3, this ${\cal N}=1$ theory has
an infinite tower of chiral primary operators of the type
$\Tr[\Phi_1^{n p}\Phi_2^{nq}]$ with $(p,q)$ being nonnegative
integers. Importantly, these operators are not chiral primaries in the
${\cal N}=4$ theory. However, it is fairly clear what happens to them as
$\beta\rightarrow 0$ or $n\to \infty$; they get infinitely large energies or
scaling dimensions. In fact all the operators of the type
$\Tr[\Phi_1^{J_1}\Phi_2^{J_2}]$ in ${\cal N}=4$ theory can be
accomodated within the tower of operators between the identity and
$\Tr[\Phi_1^n\Phi_2^n]$ as $n\rightarrow\infty$.
The operators of interest to us will satisfy the
condition 
\be
\beta J_{1,2}\ll1
\ee
which, for $\beta=1/n$ simply means that $J_{1,2}\ll n$ and so these
operators are actually closer to the bottom of the tower, and their
anomalous dimensions will get small $\beta$ -dependent
corrections. Correspondingly, in the limit of large $J_{1,2}$, the
dual classical string states will be 
small $\beta$-deformations of the two spin folded string solution of
\cite{Frolov:2003xy}. To be precise we are taking the $n\to \infty$
limit first and subsequently the limit $J_{1,2}\to\infty$.
The authors of \cite{Frolov:2005ty} 
considered operators for which $\beta J_{1,2}$ was
kept fixed in the large $J_{1,2}$ limit. 
While it is not $a$ $priori$ clear
that the two limits describe an overlapping set of states, our results
indicate that they are indeed compatible.
In fact, the agreement that we will find is implied by the
general matching  shown in \cite{Frolov:2005ty} between the string sigma model 
and the continuum limit of the coherent state action for the spin-chain.

%

In this article we calculate the energy for the $\beta$ deformed
folded string solutions and present the transcendental equations
which define the parameters classifying them.
We then perform an explicit twisted spin chain analysis in the
thermodynamical limit for this system, which should be regarded as a
generalization of the symmetrical two cuts solutions associated with
the undeformed theory. For the spin-chain/spinning string
correspondence to go through, 
we demonstrate how the elliptic moduli in
the twisted spin chain and the parameters in the folded strings should 
be naturally identified. Moreover, we calculated the anomalous
dimension using the twisted spin chain, and showed how this can be
matched with the energy of the folded strings after some elliptic
modular transformations and the identification of the moduli stated
earlier.

In the Section 2, we shall briefly describe the exactly marginal
deformations of ${\mathcal{N}}=4$ SYM, and explain how the
corresponding dual geometry was obtained. In section 3, we shall
present a discussion on the class of operators we are interested in.
Section 4 will be devoted to the folded string solution in the
Lunin-Maldacena background. The general two-cut twisted spin-chain
analysis is presented in section 5. We will conclude in section 6.
In Appendices A and B we list some useful identities for elliptic
functions and present the calculational details for the energy of the
semiclassical strings. 
 
\section{The ${\mathcal{\beta}}$ deformation of ${\mathcal{N}}=4$ SYM
  and its dual background} 

The ${\mathcal{N}}=4$ SYM theory with $SU(N)$ gauge group in
four dimensions appears as the low energy world volume theory for
a stack of $N$ coincident D3-branes. In the ${\mathcal{N}}=1$
language, this theory contains one vector multiplet and three
adjoint chiral multiplets  $\Phi_{i}$, $i=1,2,3$ and a
superpotential of the form 
\be\label{eq:be1}
W=\Tr(\Phi_{1}[\Phi_{2},\Phi_{3}])\,.
\ee
One should note that the theory, when written in the language of
${\mathcal{N}}=1$ SUSY, manifestly displays only an $SU(3)\times U(1)_R$
subgroup of the full $SU(4)$ R-symmetry of ${\cal N}=4$ SYM. The
complexified gauge coupling of the theory
$\tau=\frac{4\pi i}{g_{YM}^2}+\frac{\theta}{2\pi}$, is an exactly
marginal coupling parametrising a family of four dimensional field
theories with sixteen supercharges. The $SL(2,{\mathbb Z})$
electric-magnetic duality group acts on $\tau$ and relates different
theories on this fixed line.
%

In addition to the gauge coupling $\tau$, the ${\cal N}=4$ theory
possesses two other ${\cal N}=1$ SUSY preserving exactly marginal
deformations namely,
${\mathcal{O}}_{1}=h_{1}\Tr(\Phi_{1}\{\Phi_{2},\Phi_{3}\})$ and 
${\mathcal{O}}_{2}=h_{2}\Tr(\Phi_{1}^3+\Phi_{2}^3+\Phi_{3}^{3})$.
Arguments for their  
exact marginality were given by Leigh and Strassler 
\cite{Leigh:1995ep}, and they have been extensively studied by
earlier authors as well \cite{early}. 
The two marginal couplings $h_{1}$ and $h_{2}$, along with the gauge
coupling $\tau$ now parametrise a three (complex) parameter family of
${\mathcal{N}}=1$ superconformal field theories.  

In this article we restrict attention to a subset of these,
the so-called $\beta$ deformations obtained by setting
$h_{2}$ to zero so that up to a field 
rescaling the superpotential of the resultant
${\mathcal{N}}=1$ theory can be expressed as
\be\label{eq:be2}
W=\kappa\Tr(e^{i\pi\beta}\Phi_{1}\Phi_{2}\Phi_{3}-
e^{-i\pi\beta}\Phi_{1}\Phi_{3}\Phi_{2})\,.  
\ee 
Both $\kappa$ and $\beta$ will be complex in general. However, in this
paper we will 
only be interested in the case of real $\beta$. We can recover the
${\mathcal{N}}=4$ theory by setting $\beta=0$ and $\kappa=1$. The
$\beta$ deformation preserves a $U(1)^3$ global symmetry generated by the
Cartan subalgebra of  the original $SU(4)$ R-symmetry of
${\mathcal{N}}=4$ SYM. Put differently, 
the $SU(3) \times U(1)_{R}\subset SU(4)_R$ global symmetry of ${\cal
  N}=4$ SYM which is 
manifest in the ${\cal N}=1$ language, is broken by the $\beta$
deformation which preserves a $U(1)\times U(1)\times U(1)_R \subset SU(3)
\times U(1)_{R}$ global symmetry.  It is also worth noting
that the $SL(2,\mathbb{Z})$ invariance of ${\mathcal{N}}=4$ SYM
extends to the $\beta$ deformed theory \cite{Dorey:2002pq} via the
following action on the couplings
\be\label{eq:be3} \tau\rightarrow \frac{a \tau
+b}{c \tau +d}\,,\;\;\; \beta\rightarrow \frac{\beta}{c\tau
+d}\,,\;\;\; \left(
\begin{array}{cc}
a & b \\
c & d \end{array} \right)\in SL(2,\mathbb{Z})\,.
\ee
Various aspects of the  $\beta$ deformed theories have been
extensively studied in
\cite{Dorey:2003pp}-\cite{Berenstein:2000ux}.  

The $\beta$ deformed family of superconformal theories with $U(N)$
gauge group are also known to have string theory/supergravity duals at
large $N$ and $g^2_{YM}N>>1$. For suitably small $\beta$ the gravity dual can
be thought of as a smooth deformation of $AdS_5\times S^5$ and the
fact that the deformation is marginal implies that it should only
result in a deformation of the $S^5$, breaking the $SO(6)_R$ global
symmetry to $U(1)\times U(1)\times U(1)_R$ while preserving the
$AdS_5$ metric. In \cite{Aharony:2002hx} it was shown to lowest order
in $\beta$ that the deformation
corresponds to switching on a source for the complexified IIB
three form field strength $G_{3}=F_{3}-\tau_{s}H_{3}$ in the
$S^{5}$ of $AdS_{5}\times S^{5}$ and that
this flux back reacts and smoothly deforms the $S^{5}$. 

The
authors of \cite{Lunin:2005jy} however found the complete SUGRA
background dual to the field theory. 
In fact they demonstrated how to obtain exact SUGRA solutions for deformations
that preserve a $U(1)\times U(1)$ global symmetry. In particular the
geometry has two isometries associated to the two $U(1)$ global
symmetries and thus contains a two torus. 
As the global symmetry of the $\beta$ deformed theory is a subset of the
${\mathcal{N}}=4$ R-symmetry group, this two torus is
contained in the $S^{5}$ of the original $AdS_{5}\times S^{5}$
background and survives the deformation. At the classical level,
the $SL(2,\mathbb{R})$ isometry of the torus allows us to generate
the new ($\beta$-deformed) supergravity background by considering its
action on the 
parameter \be\label{eq:be4}
\tau_{tor}=B_{12}+i\sqrt{\det[g]}\,,\ee where $B_{12}$ is the
component of NS two form along the torus and $\sqrt{\det[g]}$ is
the volume of the torus. Note that $\tau_{tor}$ is distinct from the
complexified string/gauge theory coupling constant $\tau$. Note also
that $B_{12}$ is absent in the  
$AdS_{5}\times S^{5}$ background we started with, and the relevant
element of $SL(2,\mathbb{R})$ which yields the $\beta$ deformation acts
on $\tau_{tor}$ as \be\label{eq:be5} \tau_{tor}\rightarrow
\frac{\tau_{tor}}{1+\beta\, \tau_{tor}}\,.\ee 
In the newly
generated background, we now have a non-vanishing, non-constant
$B_{12}$ turned on. Alternatively, we can decompose the action of
(\ref{eq:be5}) and interpret it as performing a T-duality along one of
the circles of the torus, shifting the coordinate, followed by another
T-duality transformation (TsT transformation
\cite{Frolov:2005dj}). After applying these steps to the original
$AdS_{5}\times S^{5}$ with radius of curvature $R$, 
the dual supergravity solution for the $\beta$ deformed theory in the
string frame is \bea\label{eq:be6} 
ds^{2}_{str}&=&
R^{2}_{AdS}\left[ds_{AdS_{5}}^{2}+\sum_{i=1}^{3}\left(d\mu_{i}^{2}+
G\mu_{i}^{2}d\phi_{i}^{2}\right) 
+\hat{\beta}^{2}G\mu_{1}^{2}\mu_{2}^{2}\mu_{3}^{2}
\left(\sum_{i=1}^{3}d\phi_{i}\right)^{2}\right]\,,\nonumber\\ 
G&=&\frac{1}{1+\hat{\beta}^{2}(\mu_{1}^{2}\mu_{2}^{2}+
\mu_{2}^{2}\mu_{3}^{2}+\mu_{1}^{2}\mu_{3}^{2})}\,,\;\;\;\;\;
\hat{\beta}=R^{2}_{AdS}\;\beta\,,\;\;\;\;\;\sum_{i=1}^{3}\mu_{i}^{2}=1\,,\nonumber\\
R^{4}_{AdS}&=&4\pi e^{\phi_{0}}N \,,\;\;\;\;\;  e^{2\phi}=
e^{2\phi_{0}}G\,, \nonumber\\
B^{NS}&=&
\hat{\beta}R^{2}_{AdS}\;G\left(\mu_{1}^{2}\mu_{2}^{2}d\phi_{1}d\phi_{2}
+\mu_{2}^{2}\mu_{3}^{2}d\phi_{2}d\phi_{3}+\mu_{1}^{2}\mu_{3}^{2}
d\phi_{1}d\phi_{3}\right)\,,\nonumber\\
C_{2}&=&-R^{2}_{AdS}\;e^{-\phi_{0}}\hat{\beta}\,\omega_{1}d\psi\,,\;\;\;\;\;
d\omega_{1}=12\cos\alpha\sin^{3}\alpha\sin\theta\cos\theta\,
d\alpha\, d\theta\,,\nonumber\\
F_{5}&=&dC_{4}=4R^{4}_{AdS}\;e^{-\phi_{0}}(\omega_{AdS_{5}}+\omega_{S^{5}})\,.
\eea
The $AdS$ radius $R_{AdS}$ is measured in units of the string scale
$\sqrt{\alpha'}$.   
Here $\omega_{AdS_{5}}$ and $\omega_{S^{5}}$ are the volume
forms for $AdS_{5}$ and $S^{5}$ respectively, $\phi$ and
$\phi_{0}$ are the dilaton and its expectation value, $C_{2}$ and
$C_{4}$ are the RR two-form and four-form. $\mu_{1}$, $\mu_{2}$
provide the standard parametrization of $S^2$:
\be
\mu_{1}=\sin\alpha\cos\theta\,,\;\;\;\;\;\
\mu_{2}=\sin\alpha\sin\theta\;\;\;\;\;\mu_{3}=\cos\alpha\,.
\ee 
The angular variable $\psi$ is a combination of the three toroidal directions
$\psi=\frac{1}{3}(\phi_{1}+\phi_{2}+\phi_{3})$, and it is related
to the $U(1)_{R}$ generator of the ${\mathcal{N}}=1$ $SU(2,2|1)$
superconformal group. We can see that both $B^{NS}$ and $C_{2}$
are non-constant, therefore the associated non-trivial field strength
should deform the original $S^{5}$. Topologically, the compact
manifold of the solution is still $S^{5}$, and we can continuously
deform it back into $S_{5}$ by decreasing $\beta$, 
hence switching off the deformation parameters. We expect that for the
macroscopic semiclassical strings moving in $S^{5}$, we can also
continuously deform them into their counterparts in the new
background. Solutions of this kind will be the focus of this paper.

The classical supergravity solution (\ref{eq:be6}) can only be
valid if the string length is much smaller than the typical size
of the torus, therefore we have to supplement the usual condition
$R_{AdS}\gg 1$ with $R_{AdS}\;\beta\ll 1$.

\section{Operators in $\beta$ deformed theory}

In ${\mathcal{N}}=4$ SYM, physical states correspond to local gauge
invariant operators which transform in unitary
representations of the superconformal group $SU(2,2|4)$, specified
by the Dynkin labels $[\Delta,s_{1},s_{2},r_{1},r_{2},r_{3}]$ of
its bosonic subgroup $SO(1,1)\times SO(1,3)\times SU(4)_{R}$.
Here $\Delta$ is the scaling dimension (the eigenvalue of the
dilatation operator), $s_{1}$ and $s_{2}$ are the
two Lorentz spins of $SO(1,3)$, whereas $r_{1}$, $r_{2}$ and
$r_{3}$ give the Dynkin labels of $SU(4)_{R}$. Chiral
primary operators or ${\frac{1}{2}}$-BPS states of the ${\cal N}=4$ theory
are those whose scaling dimensions $\Delta$ are uniquely determined
by their R-symmetry representations. The lowest (bosonic) components
of these multiplets transform in a representation of weight $[0,k,0]$
of $SU(4)_R$.

Generic gauge invariant, single trace operators correspond to closed
string states in the Type IIB theory on $AdS_5\times S^5$. In
particular, it is well-known that single trace gauge theory operators
with large R-charges have a dual description on the string sigma model
side in terms of 
semiclassical string states 
\cite{Berenstein:2002jq,Gubser:2002tv,Frolov:2002av} carrying large
angular momentum on the $S^5$ of $AdS_5\times S^5$. We will only be
interested in those states which move on the compact manifold,
appearing as 
point particles in $AdS_5$ and for which $s_{1}$ and
$s_{2}$ can be set to zero.  

Our focus will specifically be on semiclassical string solutions
in the $\beta$ deformed theory,  corresponding to a class of operators with
$[r_{1},r_{2},r_{3}]=[J_{2},J_{1}-J_{2},J_{2}]$, {\it i.e.} the single
trace operators of the form
$\Tr(\Phi_{1}^{J_{1}}\Phi_{2}^{J_{2}})+\text{Permutations}$. The
labels here are the charges under the unbroken
$U(1)^3\subset SU(4)_R$ global symmetry in the presence of the $\beta$
deformation. For generic $\beta$, operators with charges $[0,k,0]$ (or
$[k,0,0]$ and $[0,0,k]$) are chiral primary operators
\cite{Berenstein:2000hy, Berenstein:2000ux} just as they
would have been in the ${\cal N}=4$ theory.

Both $J_1$ and $J_2$ will be taken to be large so that 
unlike typical BMN operators \cite{Berenstein:2002jq}, they will have
very high density of 
impurities and we can no longer treat these states as small deviations
from 
BPS operators. While the bare scaling dimension is equal to $J_{1}+J_{2}$,
it receives quantum corrections at all orders in the 't Hooft coupling
$\lambda=g_{YM}^{2}N$.  In particular, the problem of finding the one loop 
anomalous dimension boils down to diagonalizing a complicated mixing
matrix between single trace operators formed from various
inequivalent permutations of a string of $J_1$ $\Phi_{1}$'s and $J_2$
$\Phi_{2}$'s. For the ${\cal N}=4$ theory $(\beta=0)$ this problem was
explicitly  
solved  in \cite{Beisert:2003xu} using the Bethe ansatz, and was later
shown to precisely match the string sigma model prediction in
\cite{Frolov:2003xy,Beisert:2003ea}. 

We can generally
express the scaling dimension for this class of operators in terms of
the parameters in the theory as
$\Delta=\Delta[J,\frac{\lambda}{J^{2}},\frac{J_{1}}{J}]$, where
$J=J_{1}+J_{2}$. In the presence of the $\beta$ deformation, one 
expects the scaling dimensions of this class of operators to
receive further $\beta$ dependent corrections. These corrections would
also appear in the energy of the corresponding semiclassical
string states moving in the dual SUGRA background. The aim of this
paper is to show the matching between the anomalous dimension of the
gauge theory operators above and the semiclassical string energy in
the $\beta$ deformed theory, for a specific range of values of
$\beta$.  

Let us now specify the range of $\beta$ that we are interested
in. From the field theory perspective it is particularly interesting
to consider theories with $\beta=\frac{1}{n}$, where $n$ is a
positive integer. 
It is known that for $\beta=1/n$ and $n$ a finite integer, the $U(N)$ field
theory is in fact the low energy world volume description of a stack
of $N$ coincident D3 branes 
placed at the singularity of the orbifold
${\mathbb{C}}^{3}/({\mathbb{Z}}_{n}\times {\mathbb{Z}}_{n})$ with
discrete torsion \cite{Douglas:1998xa,Douglas:1999hq}. In the large
$N$ limit, keeping $n$ small, we can
take a near horizon limit to obtain the gravity dual description,
namely IIB string 
theory propagating in $AdS_{5}\times S^{5}/({\mathbb{Z}}_{n}\times
{\mathbb{Z}}_{n})$ \cite{Berenstein:2000hy, Berenstein:2000ux}. In
\cite{Berenstein:2000ux} the conditions for the single trace
operators to be $\frac{1}{2}$ BPS in this theory are specified.
Consider a generic single trace operator labelled
$(J_{1},J_{2},J_{3})$ \be\label{eq:dis1}
\Tr(\Phi_{1}^{J_{1}}\Phi_{2}^{J_{2}}\Phi_{3}^{J_{3}})\,. \ee 
This operator is $\frac{1}{2}$ BPS if 
\bea\label{eq:dis2}
(J_{1},J_{2},J_{3})=(k,0,0)\,,\;\;\;\;\; (0,k,0)\,,\;\;\;\;\;
(0,0,k)\quad k\in\mathbb{N}
\eea
just as it would be in the ${\cal N}=4$ theory. Remarkably however,
the ${\cal N}=1$ superconformal $\beta$ deformation with $\beta=1/n$
has an infinite set of additional chiral primary operators labelled by
\be\label{eq:dis3}
(J_1, J_2, J_3)= (k_1, k_2, k_3)\qquad k_i=0\;{\rm mod}(n).
\ee

Eq. (\ref{eq:dis2}) is simply inherited from the ${\mathcal{N}}=4$
classification while we can understand Eq. (\ref{eq:dis3}) by noticing
that using the F-term constraints, this class of operators
actually lies in the centre of the algebra of the ${\cal N}=1$
holomorphic operators and hence commutes with the superpotential. 
As a result the scaling dimensions of these operators do not get quantum
corrections. 

By this classification, a generic two spin operator of the kind we are
interested in, 
$(J_{1},J_{2},0)$,  would fall between two $\frac{1}{2}$ BPS operators
labelled by $(np,nq,0)$ and $(n(p+1),n(q+1),0)$, where $p,q$ are
non-negative integers. It is interesting to ask what happens to this
infinite tower of $1/2$ BPS operators when
$n\rightarrow\infty$, since in this limit we should recover the ${\cal
N}=4$ theory which has only the BPS states (\ref{eq:dis2}). 
When $n\to\infty$, the orbifold description
can no longer be valid, as the cone constructed from the
${\mathbb{C}}^{3}/({\mathbb{Z}}_{n}\times{{\mathbb{Z}}}_{n})$
orbifold now shrinks to a cylinder of size smaller than the string
length \cite{Lunin:2005jy,Adams:2001ne}. In fact in this limit
$\beta =1/n\ll 1$ it is appropriate to describe theory in terms of the
smooth SUGRA geometry (\ref{eq:be6}) wherein the states with labels
$(n p, n q , 0)$ with $p,q\neq 0$ will have very large masses and will
decouple. 

In this paper we will be interested in states of the $\beta$ deformed
theory which can be understood as small deformations of the two spin
states studied in \cite{Frolov:2003xy} namely the folded closed string
solution. These non-BPS operators lie in between $(0,0,0)$ and
$(n,n,0)$ states of the theory with $\beta=1/n\ll 1$, and we can
effectively fit all two spin ${\cal N}=1$ holomorphic operators in 
that range. Specifically, we are interested in the limit $\beta
J_{1},\beta J_{2}\ll 1$ with $J_{1},J_{2}\gg 1$. Since $\beta=1/n$,
these operators are actually lying close to the bottom of the tower of
states between $(0,0,0)$ and $(n,n,0)$ and can be thought of as the
$\beta$ deformed versions of the corresponding ${\cal N}=4$ states. 
We should point out that this limit is 
consistent with the requirement $R_{AdS}\;\beta\ll 1$ for the supergravity
solution $(\ref{eq:be6})$ to be valid and enables us to find the
semiclassical string solution.

Note also that the limit we are considering is 
different to that studied in \cite{Frolov:2005ty} where $\beta
J_{1,2}$ was kept fixed in the limit of large $J_1$ and $J_2$. The
latter limit will, in principle, allow the study of all states
including those that live in the middle of the
abovementioned tower of states in the $\beta$ deformed theory with
$\beta=1/n$. 
 
\section{The folded closed string in Lunin-Maldacena background}

In this section we will describe certain semiclassical solutions which are
the $\beta$ deformed versions of the folded closed string solutions of
\cite{Frolov:2003xy}.
\subsection{General equations of motion:} 
Let us first use the dual supergravity solution for the $\beta$ deformed
theory in (\ref{eq:be6}) to write down the bosonic part of string
sigma model Lagrangian 
\bea\label{eq1:fold}
S_{\rm bosonic}&=&-\frac{\sqrt{\lambda}}{4\pi}\int\,d^{2}\sigma\,
\{G^{(AdS_{5})}_{mn}\del_{a}X^{m}\del^{a}X^{n}\nonumber\\  
  &+&\sum_{i}\left(\del_{a}\mu_{i}\del^{a}\mu_{i}+G\mu_{i}^{2}
\del_{a}\phi_{i}\del^{a}\phi_{i}\right) 
+\hat{\beta}^{2}G\mu_{1}^{2}\mu_{2}^{2}\mu_{3}^{2}
(\sum_{i}\del_{a}\phi_{i})(\sum_{j}\del^{a}\phi_{j})\nonumber\\ 
&+&2\hat{\beta}G\epsilon^{a b}
\left(\mu_{1}^{2}\mu_{2}^{2}\del_{a}
\phi_{1}\del_{b}\phi_{2}+\mu_{2}^{2}\mu_{3}^{2}\del_{a}\phi_{2}
\del_{b}\phi_{3} +\mu_{1}^{2}\mu_{3}^{2}
\del_{a}\phi_{1}\del_{b}\phi_{3}\right)\}\,.  
\eea
where $\sqrt \lambda=R^2_{AdS}/\alpha'$.
As we are interested in a perturbation of the two spin solution
obtained in the 
${\mathcal{N}}=4$ case \cite{Frolov:2003xy} which only moves in the
compact manifold, we shall ignore the $AdS_{5}$ part for now. Treating
the spacetime embedding as worldsheet fields, we can deduce the
following equations of motion for the angular coordinates $\theta$
and $\alpha$: 
\bea\label{eq:fold2}
&&2\del_{a}\del^{a}\alpha=\sin 2\alpha\,\del_{a}\theta\del^{a}\theta
+\sum_{i}\frac{\del}{\del\alpha}(G\mu_{i}^{2})\del_{a}\phi_{i}\del^{a}\phi_{i}
+\hat{\beta}^{2}\frac{\del}{\del\alpha}(G\mu_{1}^{2}\mu_{2}^{2}\mu_{3}^{2})
(\sum_{i}\del_{a}\phi_{i})(\sum_{i}\del^{a}\phi_{i})\nonumber\\
&&+2\epsilon^{ab}\hat{\beta}\{\frac{\del}
{\del\alpha}(G\mu_{1}^{2}\mu_{2}^{2})\del_{a}\phi_{1}\del_{b}\phi_{2} 
+\frac{\del}{\del\alpha}(G\mu_{2}^{2}\mu_{3}^{2})\del_{a}\phi_{2}
\del_{b}\phi_{3} 
+\frac{\del}
{\del\alpha}(G\mu_{1}^{2}\mu_{3}^{2})
\del_{a}\phi_{1}\del_{b}\phi_{3}\}\,,\\\nonumber\\   
&&2\sin^{2}\alpha\,\del_{a}\del^{a}\theta+2\sin 2\alpha\,\del_{a}
\alpha\del^{a}\theta=\nonumber\\ 
&&\sum_{i}\frac{\del}{\del\theta}
(G\mu_{i}^{2})\del_{a}\phi_{i}\del^{a}\phi_{i} 
+\hat{\beta}^{2}\frac{\del}{\del\theta} (G\mu_{1}^{2}\mu_{2}^{2}\mu_{3}^{2})
(\sum_{i}\del_{a}\phi_{i})(\sum_{i}\del^{a}\phi_{i})\nonumber\\
&&+2\epsilon^{ab}\hat{\beta}\{\frac{\del}{\del\theta}(G\mu_{1}^{2}\mu_{2}^{2})
\del_{a}\phi_{1}\del_{b}\phi_{2} 
+\frac{\del}{\del\theta}(G\mu_{2}^{2}\mu_{3}^{2})\del_{a}\phi_{2}
\del_{b}\phi_{3} 
+\frac{\del}{\del\theta}(G\mu_{1}^{2}\mu_{3}^{2})\del_{a}\phi_{1}
\del_{b}\phi_{3}\}\,,
\eea
The equations of motion for $\phi_i$ which parametrise the three
$U(1)$ isometries of this background, yield the following conservation
laws for the worldsheet densities ${\cal J}^a$
\bea 
&&\del_{a}{\mathcal{J}}^{a}_{1}=\del_{a}
    {\mathcal{J}}^{a}_{2}=\del_{a}{\mathcal{J}}^{a}_{3}=0\,,\\ 
&&{\mathcal{J}}^{a}_{1}=G\mu_{1}^{2}\{\del^{a}\phi_{1}+\hat{\beta}
\epsilon^{ab}(\mu_{2}^{2}\del_{b}\phi_{2}-\mu_{3}^{2}\del_{b}\phi_{3})
+\hat{\beta}^{2}\mu_{1}^{2}\mu_{2}^{2}\mu_{3}^{2}(\sum_{i}
\del^{a}\phi_{i})\}\,,\\
&&{\mathcal{J}}^{a}_{2}=
G\mu_{2}^{2}\{\del^{a}\phi_{2}+\hat{\beta}\epsilon^{ab} 
(\mu_{3}^{2}\del_{b}\phi_{3}-\mu_{1}^{2}\del_{b}\phi_{1})
+\hat{\beta}^{2}\mu_{1}^{2}\mu_{2}^{2}\mu_{3}^{2}
(\sum_{i}\del^{a}\phi_{i})\}\,,\\ 
&&{\mathcal{J}}^{a}_{3}=G\mu_{3}^{2}\{\del^{a}\phi_{3}
+\hat{\beta}\epsilon^{ab}(\mu_{1}^{2}\del_{b}\phi_{1}-\mu_{2}^{2}
\del_{b}\phi_{2}) 
+\hat{\beta}^{2}
\mu_{1}^{2}\mu_{2}^{2}\mu_{3}^{2}(\sum_{i}\del^{a}\phi_{i})\}\,. 
\eea 
These conservation laws arise simply because the action only depends
on derivatives of $\phi_i$ and translations in these directions are
isometries. 
\subsection{Ansatz}
Here we are only interested in the two spin solutions which are
dual to the operator $\Tr(\Phi_{1}^{J_{1}}\Phi_{2}^{J_{2}})$.  
Therefore we first restrict the semiclassical solution to be along the axis
$\alpha=\frac{\pi}{2}$ and we can see that this is consistent with
the equation of motion for $\alpha$ (4.2) and effectively sets
${\mathcal{J}}_{3}^{a}$ zero. Clearly, from the $Z_{3}$ discrete symmetry
between $\Phi_{1}$,$\Phi_{2}$ and $\Phi_{3}$, we can obtain the two
spin solutions dual to the operators $\Tr(\Phi_{1}^{J_{1}}\Phi_{3}^{J_{3}})$ or
$\Tr(\Phi_{2}^{J_{2}}\Phi_{3}^{J_{3}})$ by choosing alternate axes,
$\theta=0$ or $\theta=\frac{\pi}{2}$ respectively.
Furthermore, we shall make following rotating string ansatz for the spacetime
embedding  
\be\label{eq8:fold}
t={\mathcal{\kappa}}\tau\,,\,\;\;\;\;\;
\theta\equiv\theta(\sigma)\,,\;\;\;\;\;
\phi_{1,2}=\omega_{1,2}\tau+h_{1,2}(\sigma),\qquad\phi_3=0.    
\ee 
Here $(\tau,\sigma)$ are the string worldsheet coordinates as usual,
and the string is spinning along the $\phi_1$ and $\phi_2$ directions
while also having a nontrivial spatial extent along $\theta, \phi_1$ and
$\phi_2$ coordinates.

Within this ansatz the worldsheet angular momentum densities ${\cal
  J}_{1}^{\tau}$ and ${\cal J}_{2}^{\tau}$ become independent of time
  and only
involve $\sigma$-dependent terms. This, along with the conservation
  laws $\partial_a {\cal J}^a_{1,2}=0$, in turn implies that the
  densities ${\cal J}^\sigma_{1,2}$ are constant and thus uniformly
  distributed along the string,
\bea\label{eq9:fold}
&&{\cal
  J}_1^\sigma=G\mu_{1}^{2}(\frac{dh_{1}}{d\sigma}+
  \hat{\beta}\mu_{2}^{2}\omega_{2})=C_{1}  
  \,,\;\;\;\;\;\qquad\quad
{\cal J}_2^\sigma=G\mu_{2}^{2}(\frac{dh_{2}}
  {d\sigma}-\hat{\beta}\mu_{1}^{2}\omega_{1})=C_{2}\,,\\ 
&&\text{with}\;\;\;\;\; \mu_{1}= \cos\theta\,,\;\;\;\;\;
\mu_{2}=\sin\theta\,,\;\;\;\;\;G
  =\frac{1}{1+\hat{\beta}^{2}\cos^{2}\theta\sin^{2}\theta}\,.\nonumber 
\eea
Here $C_{1}$ and $C_{2}$ are the integration constants. We
will see subsequently that turning on non-zero values for $C_{1}$ and
$C_{2}$ is necessary to incorporate the effect of the deformation
parameter $\beta$ into the solution.
From (\ref{eq9:fold}), the equations of motion for $h_{1}$ and $h_{2}$
can be written in terms of $\theta(\sigma)$ 
\bea\label{eq10:fold}
&&\frac{dh_{1}}{d\sigma}=\frac{C_{1}}{\mu_{1}^{2}}+\hat{\beta}\;
  \Omega_{2}\;\mu_{2}^{2}\,,\;\;\;\;\;\qquad\qquad 
\frac{dh_{2}}{d\sigma}=
  \frac{C_{2}}{\mu_{2}^{2}}-\hat{\beta}\;
  \Omega_{1}\;\mu_{1}^{2}\,,\\\nonumber\\   
&&{\Omega_2}=-(\omega_2-\hat{\beta}C_{1}),\qquad\qquad\qquad\quad\;
  \Omega_{1}=-(\omega_1+\hat{\beta}C_{2})  
\eea  
The isometries in the $\phi_{1,2}$ directions lead to a conservation
  of the corresponding angular momenta $J_1$ and $J_2$. These  are
  naturally obtained as worldsheet spatial integrals over the
  associated charge densities ${\cal J}_{1,2}^\tau$
\be\label{eq11:fold}
J_{1}=\frac{\sqrt{\lambda}}
  {2\pi}\int^{2\pi}_{0}d\sigma
  \left\{\Omega_{1}\cos^{2}\theta(\sigma)\right\}\,,\;\;\;\;\; 
J_{2}=\frac{\sqrt{\lambda}}{2\pi}\int^{2\pi}_{0}
  d\sigma\left\{\Omega_{2}\sin^{2}\theta(\sigma)\right\}\,. 
\ee
Note that these have exactly the same form as the expressions encountered 
in \cite{Frolov:2003xy} for the two spin folded closed string in
  $AdS_5\times S^5$ with the replacement $\omega_{1,2}\rightarrow
 -\Omega_{1,2}$.   
Combining (\ref{eq9:fold}) and (\ref{eq10:fold}), the
equation of motion for $\theta(\sigma)$ becomes \be\label{eq12:fold}
\frac{d^{2}\theta}{d\sigma^{2}}=
-\Omega_{12}^{2}\cos\theta\sin\theta
+C_{2}^{2}\frac{\cos\theta}
  {\sin^{3}\theta}-C^{2}_{1}\frac{\sin\theta}{\cos^{3}\theta}\,,\;\;\;\;\; 
\Omega_{12}^{2}=\Omega_{2}^{2}-\Omega_{1}^{2}\,. 
\ee 
With non-zero constants $C_1$ and $C_2$ this equation describes a
deformation of the folded closed string of \cite{Frolov:2003xy}. 
Clearly, we could also choose $\Omega_{1}=\Omega_{2}$ and $C_{1}=C_{2}=0$
resulting in a solution with fixed $\theta$ which is the so-called
``circular'' string. More generally, integrating (\ref{eq12:fold}) and
  introducing appropriate integration constants we get 
\be\label{eq13:fold}
\left(\frac{d\theta}{d\sigma}\right)^{2}=\Omega_{12}^{2}
  (\sin^{2}\theta_{0}-\sin^{2}\theta)  
+C_{1}^{2}\left(\frac{1}{\cos^{2}\theta_{0}}-
\frac{1}{\cos^{2}\theta}\right) +
C_{2}^{2}\left(\frac{1}{\sin^{2}\theta_{0}}-
\frac{1}{\sin^{2}\theta}\right)\,.\ee 
This equation now governs the entire dynamics of our new solution,
and in conjunction with ({\ref{eq10:fold}) yields the following form for the 
Virasoro constraints
\be\label{eq14:fold}
\kappa^{2}=\Omega_{1}^{2}\cos^{2}\theta_{0}+\Omega_{2}^{2}\sin^{2}\theta_{0}
+(\frac{C_{1}^{2}}{\cos^{2}\theta_{0}}+\frac{C_{2}^{2}}{\sin^{2}\theta_{0}})\,,
\ee \be\label{eq15:fold}
\omega_{1}C_{1}+\omega_{2}C_{2}=-(\Omega_{1}C_{1}+\Omega_{2}C_{2})=0\,.
\ee
The first Virasoro constraint gives the energy of the 
string configuration 
\be\label{eq:energydef}
E^2 =\lambda\kappa^2. 
\ee
The first two terms in (\ref{eq14:fold}) together constitute the 
energy of the original two spin folded string, 
while the last term can be thought of as a correction 
due to non-zero $\beta$ in the new background. 
We will see below that the second Virasoro constraint
(\ref{eq15:fold}) actually implies a relation between the two
integration constants $C_{1}$ and $C_{2}$. Equations (4.14) and (4.15)
have also appeared in the two spin solutions 
for the so-called ``Neumann-Rosochatius'' integrable system discussed
in \cite{Arutyunov:2003uj} as a generalization of the
semiclassical strings in the undeformed $AdS_{5}\times S^{5}$.
\subsection{Solutions in Elliptic parametrization}
Let us now solve (\ref{eq13:fold}) in the elliptic
parametrization. We first introduce a new
variable $x=\sin^{2}\theta$ so that the equation can then be rewritten as
\bea\label{eq16:fold}
&&\left(\frac{dx}{d\sigma}\right)^{2}=
4\Omega_{12}^{2}(x_{+}-x)(x_{0}-x)(x-x_{-}),
\eea
where $x_{0}=\sin^{2}\theta_{0}\;$ and 
\bea
&&x_{\pm}=\\\nonumber\\\nonumber
&&\frac{1}{2}\left[1+\left(\frac{C_{1}^{2}}{(1-x_{0})\Omega_{12}^{2}}
+\frac{C_{2}^{2}}{x_{0}\Omega_{12}^{2}}\right)\pm
\sqrt{
\left(1+\frac{C_{1}^{2}}{(1-x_{0})\Omega_{12}^{2}}
+\frac{C_{2}^{2}}{x_{0}\Omega_{12}^{2}}\right)^{2}
-4\frac{C_{2}^{2}}{x_{0}\Omega_{12}^{2}}}\right]
\eea

or alternatively
\be
\left(\frac{C_{1}}{\Omega_{12}}\right)^{2}=(x_{+}-1)(1-x_{0})(1-x_{-})\,,
\;\;\;\;\; \left(\frac{C_{2}}{\Omega_{21}}\right)^{2}=x_{+}x_{0}x_{-}\,.
\ee
Using the periodicity of the closed string fields in the $\sigma$
coordinate, we can 
integrate both sides of (\ref{eq16:fold}) and express the results  
as complete elliptic
integrals (see Appendix A for our conventions)
\be
2\pi=\int^{2\pi}_{0}d\sigma=\frac{2}
{\Omega_{12}}\int^{x_{0}}_{x_{-}}\frac{dx}
{\sqrt{(x_{+}-x)(x_{0}-x)(x-x_{-})}}\nonumber
\ee 
leading to 
\be\label{eq17:fold}
\Omega_{12}={\frac{2}{\pi}}\frac{{\mathrm{K}}(k)}{\sqrt{x_{+}-x_{-}}}\,,
\;\;\;\;\;\qquad k=\frac{x_{0}-x_{-}}{x_{+}-x_{-}}\,. 
\ee
Similarly, the angular momentum integrals of motion are
\bea\label{eq18:fold}
{J}_{1}&=&\sqrt\lambda\frac{\Omega_{1}}{2\Omega_{12}}\int^{x_{0}}_{x_{-}}
\frac{dx\,(1-x)}{\sqrt{(x_{+}-x)(x_{0}-x)(x-x_{-})}}
\nonumber\\\nonumber\\  
&=&\Omega_{1}\left(1-\left(x_{+}-(x_{+}-x_{-})\frac{{\mathrm{E}}(k)} 
{{\mathrm{K}}(k)}\right)\right)\\\nonumber\\  
{J}_{2}&=&\sqrt\lambda\frac{\Omega_{2}}
{2\Omega_{12}}\int^{x_{0}}_{x_{-}}\frac{dx\,x} 
{\sqrt{(x_{+}-x)(x_{0}-x)(x-x_{-})}}\nonumber\\\nonumber\\
&=&\Omega_{2}\left(x_{+}-(x_{+}-x_{-})\frac{{\mathrm{E}}(k)}
{{\mathrm{K}}(k)}\right)\,.
\eea 
Since the variables $k,x_{0},x_{+}$ and $x_{-}$ are related via
(\ref{eq17:fold}),  
only three of these parameters are independent. We shall use $k$ and $x_{0}$
interchangeably wherever necessary, to simplify the expressions.
Using $\Omega_{12}^2=\Omega_2^2-\Omega_1^2$ and the equations above to
eliminate the variables $\Omega_1$ and $\Omega_2$, we can derive a
transcendental  
equation relating the various elliptic parameters and the angular
momentum quantum numbers 
\be\label{eq19:fold}
\frac{4\lambda}{\pi^{2}(x_{+}-x_{-})}
=\frac{J_{2}^{2}}{\left(x_{+}
  {\mathrm{K}}(k)-(x_{+}-x_{-}){\mathrm{E}}(k)\right)^{2}} 
-\frac{J_{1}^{2}}{\left((1-x_{+})
  {\mathrm{K}}(k)+(x_{+}-x_{-}){\mathrm{E}}(k)\right)^{2}}\,. 
\ee 
In the limit of vanishing $C_{1}$ and $C_{2}$, this relation 
reduces to the corresponding equation for the original two
spin folded string in \cite{Frolov:2003xy}. For non-zero $C_{1}$ and
$C_{2}$ the situation is somewhat complicated: Firstly, the second
Virasoro constraint (\ref{eq15:fold}) can no longer be trivially
satisfied and takes the form 
\be\label{eq:vir}
\frac{J_{1}C_{1}}
     {\left((1-x_{+}){\mathrm{K}}(k)+(x_{+}-x_{-})
       {{\mathrm{E}}(k)}\right)}+
\frac{J_{2}C_{2}}
     {\left(x_{+}{\mathrm{K}}(k)-
       (x_{+}-x_{-}){{\mathrm{E}}(k)}\right)}=0\,. 
\ee 
The semiclassical string solution of the type considered here can only
exist if the integration constants $C_{1,2}$ and the angular
momenta $J_{1,2}$ satisfy the above condition. 

Moreover, the dual supergravity solution for the deformed
theory with $\beta=1/n\ll1$, given by (\ref{eq:be6}), is a smooth
background. Unlike the in \cite{Douglas:1998xa,Douglas:1999hq} 
there is no orbifold action and the closed strings can only have
zero or integer winding numbers. Inspecting the form of
${dh_{i}}/{d\sigma}$ in (\ref{eq10:fold}) and integrating with respect
to $\sigma$, the winding numbers depend on the combinations
$\beta J_{1}$ and $\beta J_{2}$ which are non-integers in
general, and in particular for us $\beta J_{1,2}\ll1$. The terms
proportional to $C_1$ and $C_2$ then become 
necessary to yield integer winding numbers. 
Explicitly, we obtain 
\bea 
2\pi N_{1}&=&\phi_{1}(2\pi)-\phi_{1}(0)=\int^{2\pi}_{0}d\sigma
\frac{dh_{1}}{d\sigma}=
\int^{2\pi}_{0}d\sigma\left(\frac{C_{1}}{\mu_{1}^{2}}+\hat{\beta}
\Omega_{2}\mu_{2}^{2}\right)\nonumber\\\nonumber\\ 
&=&{\frac{4 C_1}{\Omega_{21}}}\;\frac{{\mathrm{\Pi}}\left(\alpha_{1}^{2},k\right)}
      {(1-x_{-})\sqrt{x_{+}-x_{-}}}   
\;+\;2\pi\beta J_{2}\,,
\eea
and
\bea
2\pi N_{2}&=&\phi_{2}(2\pi)-\phi_{2}(0)=\int^{2\pi}_{0}d\sigma
\frac{dh_{2}}{d\sigma}=
\int^{2\pi}_{0}d\sigma\left(\frac{C_{2}}{\mu_{2}^{2}}-
\hat{\beta}\Omega_{1}\mu_{1}^{2}\right)\nonumber\\\nonumber\\ 
&=&{\frac{4C_{2}}{\Omega_{21}}}\frac{{\mathrm{\Pi}}\left(\alpha_{2}^{2},k\right)}
    {x_{-}\sqrt{x_{+}-x_{-}}} 
-2\pi\beta J_{1}\,,\;\;\;\;\;
\eea
where
\be
\alpha_{1}^{2}=\frac{(x_{0}-x_{-})}{(1-x_{-})}\,,
\;\;\;\;\;\alpha^{2}_{2}=-\frac{(x_{0}-x_{-})}{x_{-}}\,.
\ee 
and $\Pi(\alpha^2, k)$ is a complete elliptic integral of the third
kind. 

Here $N_{1}$ and $N_{2}$ are the integer winding numbers along the
$\phi_1$ and $\phi_2$ circles. As we are
interested in the limit where $\beta J_{1},\beta J_{2}\ll 1$ we expect  
$C_{1}$ and $C_{2}$ to be parametrically small quantities and
importantly, to give a parametrically small correction to the energy
expression (\ref{eq14:fold}).

Interestingly (\ref{eq13:fold}), (\ref{eq14:fold}), and
(\ref{eq15:fold}) are also the equations of motion and
Virasoro constraints for a folded semiclassical string
configuration with arbitrary integer winding numbers
propagating in the original $AdS_{5}\times S^{5}$ background
\cite{Arutyunov:2003za}. 
Here we derived them purely from the new
dual background as a general two spin solution for the
$\beta$ deformed theory. This observation is consistent with the
relationship between semiclassical strings in the undeformed
and $\beta$ deformed backgrounds, pointed out in \cite{Frolov:2005dj}.

As noted earlier and from (\ref{eq14:fold}) we expect that $C_{1}$ and
$C_{2}$ should be small and both should vanish as $\beta\to 0$ as our solution
should only be a small perturbation away from the original folded string
solution. Naively therefore, from our discussion above, it would seem natural
to set $N_{1}$ and $N_{2}$ to zero for the configuration
corresponding to the perturbation of the two spin solution. However,
it turns out that the $\beta\to 0$ limit is somewhat subtle. 
In fact we find that in order for a well defined $\beta\to 0$ limit
(and $C_{1},C_{2}\to 0$) to exist we must have $N_2=1$.
This arises in the $\beta,C_{1,2}\to 0$ limit due to a cancellation
between a vanishing denominator in ${1}/{x_{-}}$ and a zero of the
elliptic function $\Pi(\alpha_2^2, k)$.

\subsection{Energy of states}
One can in principle derive a double expansion in $J^{-1}$ and
$\beta J\ll1$, ($J=J_1+J_2$) for the energy of the semiclassical string
solution with string $\alpha'$ corrections suppressed by higher
powers of $1/J$. In such a scheme, the coefficient of each power of
$1/J$ would be a nontrivial function of $\beta$. In particular, the energy
$E=J+\frac{\lambda}{J}\varepsilon_{1}+O(\frac{1}{J^{2}})$ where 
$\varepsilon_{1}$ would be a function of $\beta J$. 
We would need to extract $\varepsilon_{1}$ in order to compare with
the field theoretic spin chain analysis below 
which yields the anomalous dimension at one loop order in
the 't Hooft coupling $\lambda$. In order to do this systematically we
would first need to expand all the elliptic modular parameters of the
solutions above in powers of $1/J$, 
\be
k=k^{(0)}+\frac{k^{(2)}}{{\mathcal{J}}^{2}}\dots\,,\;\;\;\;\;
x_{0}=x_{0}^{(0)}+\frac{x_{0}^{(2)}}{{\mathcal{J}}^{2}}\dots\,,\;\;\;\;\;
x_{+}=x_{+}^{(0)}+\frac{x_{+}^{(2)}}{{\mathcal{J}}^{2}}\dots\,,\;\;\;\;\;
x_{-}=x_{-}^{(0)}+\frac{x_{-}^{(2)}}{{\mathcal{J}}^{2}}\dots\,, \ee
where ${\cal J} =J/\sqrt\lambda$ and the expansion coefficients
$k^{(0)},k^{(2)},x_{0}^{(0)},x_{0}^{(2)},
x_{+}^{(0)},x_{+}^{(2)},x_{-}^{(0)}, x_{-}^{(2)}$ are functions of
$\beta J$. Substituting these into (4.20),(4.22) and (4.23), after
lengthy but straightforward expansions, one can in principle rewrite
$k^{(2)},x_{+}^{(2)},x_{-}^{(2)}$ in terms of
$k^{(0)},x_{+}^{(0)},x_{-}^{(0)}$, and derive an expression of
$\varepsilon_{1}$ involving only 
$k^{(0)},x_{+}^{(0)},x_{-}^{(0)}$. We present an alternate, 
simpler derivation for $\varepsilon_{1}$ explicitly in the Appendix B
and  the result  is given by
\be \label{eq:dim}
\varepsilon_{1}
=\frac{2{\mathrm{K}}\left(k^{(0)}\right)\left({\mathrm{E}}
\left(k^{(0)}\right)-\left(1-k^{(0)}\right)
{\mathrm{K}}\left(k^{(0)}\right)\right)}{\pi^{2}}
+\frac{2{\mathrm{K}}\left(k^{(0)}\right)^{2}
\left(x_{+}^{(0)}+x_{-}^{(0)}-1\right)}{\pi^{2}
\left(x_{+}^{(0)}-x_{-}^{(0)}\right)}
\ee 
In the limit of vanishing $\beta$ wherein $x_{+}^{(0)}\to 1$,
$x_{-}^{(0)}\to 0$, and $k^{(0)}\to x_{0}^{(0)}$, the expression for
$\varepsilon_{1}$ reduces to Eq. (2.7) in
\cite{Beisert:2003ea}. The remarkable simplicity of the expression
(\ref{eq:dim}) and the fact that in the $\beta\to 0$ limit it matches
the energy of the original folded string in $AdS_5\times S^5$, 
together indicate that the $\beta$ dependent correction to the
semiclassical string energy in the deformed background can indeed be
regarded as a small perturbation. 

The $\beta$ dependences of $k^{(0)},x_{+}^{(0)}$ and
$x_{-}^{(0)}$ are given implicitly by the following transcendental
equations 
\bea\label{eq:filling}
&&\alpha = {\frac{J_{2}}{J}}=x_{+}^{(0)}-(x_{+}^{(0)}-x_{-}^{(0)})
\frac{{\mathrm{E}}\left(k^{(0)}\right)}
     {{\mathrm{K}}\left(k^{(0)}\right)}\,,\\\nonumber\\  
&&C_{1}^{(0)}+C_{2}^{(0)}=0 \longrightarrow
\left(x_{+}^{(0)}-1\right)\left(1-x_{0}^{(0)}\right)
\left(1-x_{-}^{(0)}\right)=  
x_{+}^{(0)}x_{0}^{(0)}x_{-}^{(0)}\,,\\\nonumber\\
&&2\pi\beta J(1-2\alpha)=\nonumber\\\nonumber\\
&&4\sqrt{\frac{\left(x_{+}^{(0)}-1\right)
    \left(1-x_{0}^{(0)}\right)}{\left(1-x_{-}^{(0)}\right)
    \left(x_{+}^{(0)}-x_{-}^{(0)}\right)}}  
{\mathrm{\Pi}}\left(\alpha_{1}^{2},k^{(0)}\right)+
4\sqrt{\frac{x_{+}^{(0)}x_{0}^{(0)}}{x_{-}^{(0)}\left(x_{+}^{(0)}-x_{-}^{(0)}\right)}}
{\mathrm{\Pi}}\left(\alpha_{2}^{2},k^{(0)}\right)-2\pi\,.
\nonumber\\
\eea
where $\alpha=J_2/J$ is the filling ratio. It is from
these equations that the parameters
$k^{(0)},x_{+}^{(0)}$ and $x_{-}^{(0)}$ acquire dependences on
$\beta$. One 
can, in principle, solve these complicated simultaneous equations by
expanding in powers of $\beta J$ from a given undeformed folded string
solution. We will adopt a more direct and less messy approach.

As we will demonstrate in the next section, from results at lowest
order in the $1/J$ expansion the parameters
$k^{(0)},x_{+}^{(0)}$ and $x_{-}^{(0)}$ above can be 
related to the moduli of the elliptic curve for the corresponding
spin chain with the help of some suitable elliptic modular
transformations. Following this identification of the moduli, we are
then able to reproduce the precise expression for the one loop anomalous
dimension $\varepsilon_{1}$ above from the spin-chain analysis,
provided we impose an extra 
condition, which happily turns out to be the Virasoro
constraint for the spinning string solution. 

\section{The Twisted Spin-Chain analysis}
\subsection{The Bethe ansatz}
In this section, we carry out the spin-chain analysis for the
one loop anomalous dimension of the operators
$\Tr[\Phi_1^{J_1}\Phi_2^{J_2}]$ in the $\beta$ deformed theory. As
established in \cite{Frolov:2005ty, Berenstein:2004ys}, the effect of
turning on a real valued $\beta$ deformation of the ${\mathcal{N}}=4$ SYM
Lagrangian is to introduce a non-trivial ``twisting'' parameter in
the original XXX $SU(2)$ spin chain which computes the one loop
anomalous dimensions of the single trace gauge invariant operators.
Explicitly, in the presence of the twisting 
the modifield Bethe equations are
\be
e^{-2i\pi\beta
J}\left(\frac{u_{k}+\frac{i}{2}}{u_{k}-\frac{i}{2}}\right)^{J}=\prod_{j\neq
k=1}^{J_{2}}\frac{u_{k}-u_{j}+i}{u_{k}-u_{j}-i}\,, 
\ee 
and 
\be\label{eq2:bethe}
e^{-2i\pi\beta
J_{2}}\prod^{J_{2}}_{j=1}\frac{u_{j}+\frac{i}{2}}{u_{j}-\frac{i}{2}}=1\,.
\ee 
The equation \ref{eq2:bethe} corresponds to the modification of the
trace condition due to $\beta$. One can perhaps most easily observe
this by using the Coulomb branch condition and setting $\beta$ to a
rational number. This gives a linear shift in the the total lattice
momentum of the spin chain. As we are interested in the so-called
``thermodynamic'' limit where $J,J_{2}\to \infty$ while keeping the
filling ratio $\alpha=\frac{J_{2}}{J}$ fixed, all the Bethe roots
$u_{k}$ are of order $J$. Hence we can rescale the roots,
$u_{k}=Jx_{k}$ and in terms of the rescaled Bethe roots the 
expression for the energy takes the form 
\be\label{eq:energy}
\gamma=\frac{\lambda}{8\pi^{2}J^{2}}
\sum^{J_{2}}_{k=1}\frac{1}{x_{k}^{2}}\,. 
\ee 
The twisted thermodynamical Bethe equation gets modified into
\be\label{eq:bethe} 
\frac{1}{x_{k}}=2\pi(\beta J-n_{k})+\frac{2}{J}\sum_{j\neq
k=1}^{J_{2}}\frac{1}{(x_{k}-x_{j})}\,.
\ee 
Summing over the index $k$ on both sides of the
Bethe equation yields the condition 
\be\label{eq4:bethe}
\sum_{k=1}^{J_{2}}\frac{1}{x_{k}}=2\pi((\beta
J)J_{2}-\sum_{k=1}^{J_{2}}n_{k})\,. 
\ee 

We first recall that in the absence of $\beta$ the spin chain
configuration for the folded string does not have any lattice
momentum. Then, comparing equation ({\ref{eq4:bethe}}) with the 
logarithm of (\ref{eq2:bethe}) in the thermodynamic limit, 
we deduce 
\be
\sum_{k=1}^{J_{2}}n_{k}=0. 
\ee
Moreover, since we are looking for the
lowest energy distribution of the Bethe roots, from the energy
(\ref{eq:energy}) we infer that for $J_2$ an even number, the 
$n_{k}$ should come in
pairs taking the values $\pm 1$. This means that to a first approximation,
the Bethe roots condense into two cuts labeled by
${\mathcal{C}}_{\pm}$ spread around $n_{k}=\pm 1$ respectively.
(While the condition $\sum_{k=1}^{J_{2}}n_{k}=0$ can also be
satisfied by choosing other set of $n_{k}$, these would correspond
to the multi-cut distributions and would have higher energy
$\gamma$.) 

In the undeformed theory this was essentially the complete
story - the two cuts were symmetrical about the imaginary axis with
the Bethe roots repelling one another via Coulomb-like forces on each
cut. In the undeformed case the entire distribution could be described
by the end-points of either
one of the two cuts, as the branch cuts are simply mirror
images of each other. 

Turning on a $\beta\neq 0$ gives different shifts to the
roots on the ${\mathcal{C}}_{+}$ and ${\mathcal{C}}_{-}$ cuts and as
a result, breaks the symmetry about the imaginary axis. This
asymmetry is clearly exhibited in the Bethe equation
(\ref{eq:bethe}). The four end-points of the branch cuts must now be
treated independently, although their locations are only small
perturbations around the symmetric $\beta=0$ case. This then requires us to
perform a general two-cut analysis. The appearance of general two-cut
distributions is not surprising, given the appearance of
elliptic functions describing the semiclassical string solutions
in the previous section. The Riemann surface associated to a two-cut
distribution is a torus  
on which elliptic functions are naturally defined as analytic functions.

\subsection{The two-cut Riemann surface}
Let us now solve for this two cut distribution by first labelling the four
end points as $x_{1}>x_{2}>x_{3}>x_{4}$. The two
cuts are given by ${\mathcal{C}}^{+}\in [x_{1},x_{2}]$ and
${\mathcal{C}}^{-}\in [x_{3},x_{4}]$. Our analysis will follow the
one in \cite{Kazakov:2004qf} for the untwisted spin chain.

Define the density and the resolvent of the Bethe roots in the usual
way \be
\rho(y)=\frac{1}{J}\sum_{k=1}^{J_{2}}\delta(y-x_{k})\,,\;\;\;\;\;
G(x)=\frac{1}{J}\sum_{k=1}^{J_{2}}\frac{1}{x-x_{k}}=
\int_{{\mathcal{C}}}\frac{dy\,\rho(y)}{x-y}\,,\ee 
where ${\mathcal{C}}={\mathcal{C}}_{+}\bigcup{\mathcal{C}}_{-}$. The
resolvent $G(x)$ is an analytic function with branch cuts at ${\cal
  C}^{\pm}$ and its discontinuity across a cut gives the density
of Bethe roots.

The filling fraction of the Bethe roots on each cut can be defined
as 
\be\label{eq:filling2} 
{\mathcal{S}}_{\pm}=\int_{{\mathcal{C}}_{\pm}}dx\,
\rho(x)=\frac{1}{2\pi i}\oint_{{\mathcal{C}}_{\pm}}dx\,G(x)\,,
\;\;\;\;\; {\mathcal{S}}_{+}+{\mathcal{S}}_{-}=\alpha\,, 
\ee 
with $\alpha$ given by $J_{2}/J$. The
resolvent function $G(x)$ allows us to rewrite the the momentum and
energy conditions elegantly as contour integrals
\bea 
&&2\pi \alpha(\beta J)=\int_{{\mathcal{C}}}dx
\frac{\rho(x)}{x}
=\frac{i}{2\pi}\oint_{{\mathcal{C}}}dx\frac{G(x)}{x}\,,
\\\nonumber\\\label{eq:gamma}
&&\gamma=\frac{\lambda}{8\pi^{2}J}\int_{{\mathcal{C}}}dx
\frac{\rho(x)}{x^{2}}=\frac{i\lambda}{16\pi^{3}J}
\oint_{{\mathcal{C}}}dx\frac{G(x)}{x^{2}}\,.
\eea
The equalities above can be proven by
exchanging the order of integrations and deforming the contour. Note that
(\ref{eq:filling2}) also implies ${\mathcal{S}}_{+}={\mathcal{S}}_{-}$
which can be proven by a contour deformation argument. In
our case this is consistent with the earlier cyclicity condition
$\sum_{k=1}^{J_{2}}n_{k}=0$. 

Following \cite{Kazakov:2004qf}, we now define the so-called
``quasi-momentum'' to be 
\be p(x)=G(x)-\frac{1}{2x}+\pi(\beta
J)\,,
\ee 
which is a double valued function of $x$ which satisfies the ``integer
condition''
\be\label{eq:integer}
p(x+i\epsilon)+p(x-i\epsilon)=2\pi
n_{\pm}\,,\;\;\;\;\;\ x\in{\mathcal{C}}_{\pm}\,,\;\;\;\;\;
n_{\pm}=\pm 1\,.
\ee
Naively, the extra $\pi(\beta J)$ in the definition of $p(x)$
does not appear to affect the value of $\gamma$ obtained from (\ref{eq:gamma}),
as its contribution in the contour integral is a double pole which
gives vanishing residue. 
However, the parameters entering $p(x)$ namely, the locations of the
branch points $x_{1},x_{2},x_{3}$ and $x_{4}$ can and do depend on
$\beta$ and hence the situation is somewhat subtle.

To proceed further, let us first describe the  Riemann surface for
$p(x)$.  It consists of two sheets 
with branch cuts at ${\mathcal{C}}_{+}$ and ${\mathcal{C}}_{-}$,
{\it i.e.} it is a torus. Following the discussion on the
quasi-momentum in \cite{Kazakov:2004qf}, the additional term
$\pi(\beta J)$ does 
not change the fact that $dp(x)$ is a meromorphic differential on
its Riemann surface with only double poles at $x=0$ on the two
sheets. We can describe this Riemann surface by a hyperelliptic
curve 
\be \Sigma\,:\;\;\;\;\;
y^{2}=\prod_{k=1}^{4}(x-x_{k})=x^{4}+f_{1}x^{3}+f_{2}x^{2}+f_{3}x+f_{4}\,,
\ee 
and the meromorphic differential is given by
\be
dp(x)=\frac{g(x)dx}{y},
\ee 
where $g(x)$ is a rational function.

The precise form of $g(x)$ can be fixed by asymptotic
behaviour at large and small $x$. 
First consider the behaviour as $x\to\infty$. In this
limit $G(x)\to\alpha/x$ and therefore $dp(x)\to
({1}/{2}-\alpha){dx}/{x^{2}}$. This condition restricts
$g(x)$ to be $\sum_{n\leq 1}a_{n}x^{n-1}$, with
$a_{1}=(\frac{1}{2}-\alpha)$. The presence of a double pole
in $dp(x)$ at $x=0$ then further truncates the expansion for
$g(x)$ to $a_{1}+\frac{a_{0}}{x}+\frac{a_{-1}}{x^{2}}$, so that 
near $x=0$ 
\be
dp(x)=\frac{dx}{y}\left((\frac{1}{2}-\alpha)+\frac{a_{0}}{x}
+\frac{a_{-1}}{x^{2}}\right)=\frac{dx}{2x^{2}}+O(1)\,.
\ee 
Matching coefficients in the above expansion, we deduce
that 
\be
a_{-1}=\frac{\sqrt{f_{4}}}{2}=
\frac{\sqrt{\prod_{k=1}^{4}x_{k}}}{2}\,,\;\;\;\;\;
a_{0}=\frac{f_{3}}{4\sqrt{f_{4}}}=
\frac{-\sqrt{\prod_{k=1}^{4}x_{k}}}{4}
\left(\sum_{k=1}^{4}\frac{1}{x_{k}}\right)\,.
\ee
Notice that for the undeformed theory where the Bethe roots condense
into symmetrical cuts, $a_{0}$ was identically zero. This, however, is
not the case for the general two-cut solution we are dealing with here.

\subsection{Elliptic parametrization}
The quasi-momentum $p(x)$ is necessarily a single valued function on 
each sheet (for our case without the condensate), 
namely the integral of the meromorphic
differential $dp(x)$ along a contour enclosing ${\mathcal{C}}_{+}$
or ${\mathcal{C}}_{-}$ vanishes \cite{Kazakov:2004qf}. 
This gives us the so-called ``A-cycle'' integration condition
\bea
&&0=\oint_{{\mathcal{A}}_{+}}dp(x)=
2i\int^{x_{1}}_{x_{2}}dx\,
\frac{a_{1}+\frac{a_{0}}{x}+\frac{a_{-1}}{x^{2}}}
{\sqrt{(x_{1}-x)(x-x_{2})(x-x_{3})(x-x_{4})}}
\nonumber\\\nonumber\\
&&=\frac{2i}{\sqrt{(x_{1}-x_{3})(x_{2}-x_{4})}}
\left\{\left[(1-2\alpha)-\frac{(x_{1}x_{2}+x_{3}x_{4})}
{2\sqrt{x_{1}x_{2}x_{3}x_{4}}}\right]{\mathrm{K}}(r)
+\frac{(x_{1}-x_{3})(x_{2}-x_{4})}{2\sqrt{x_{1}x_{2}x_{3}x_{4}}}
{\mathrm{E}}(r)\right\}\,.\nonumber\\\label{eq:acycle}
\eea
The integer condition (\ref{eq:integer}) translates into the
``B-cycle'' integration condition
\bea
&&4\pi=\oint_{{\mathcal{B}}}dp(x)=
2\int^{x_{2}}_{x_{3}}dx\,\frac{a_{1}+\frac{a_{0}}{x}+ 
\frac{a_{-1}}{x^{2}}}
{\sqrt{(x_{1}-x)(x_{2}-x)(x-x_{3})(x-x_{4})}}
\nonumber\\\nonumber\\
&&=\frac{2}{\sqrt{(x_{1}-x_{3})(x_{2}-x_{4})}}
\left\{\left[(1-2\alpha)-\frac{(x_{1}x_{4}+x_{2}x_{3})}
{2\sqrt{x_{1}x_{2}x_{3}x_{4}}}\right]{\mathrm{K}}(r')
-\frac{(x_{1}-x_{3})(x_{2}-x_{4})}
{2\sqrt{x_{1}x_{2}x_{3}x_{4}}}
{\mathrm{E}}(r')\right\}\,.\nonumber\\
\eea 
The elliptic modulus $r$ and its complement $r'=1-r$ are given
by the cross ratios 
\be
r=\frac{(x_{1}-x_{2})(x_{3}-x_{4})}{(x_{1}-x_{3})(x_{2}-x_{4})}\,,\;\;\;\;\;
r'=\frac{(x_{1}-x_{4})(x_{2}-x_{3})}{(x_{1}-x_{3})(x_{2}-x_{4})}\,.
\ee 
From the A and B-cycle conditions and using the Legendre relation (A.6) 
we rewrite ${\mathrm{K}}(r)$ as
\be\label{eq:kr}
{\mathrm{K}}(r)=-\frac{\sqrt{(x_{1}-x_{3})(x_{2}-x_{4})}}
{8\sqrt{x_{1}x_{2}x_{3}x_{4}}}\,.
\ee 
We shall use this result later.

The filling ratios ${\mathcal{S}}_{\pm}$ can also be explicitly
computed from the contour integrals of $p(x)$ enclosing each cut, 
\bea
{\mathcal{S}}_{+}&=&\frac{1}{\pi}\int_{x_{2}}^{x_{1}}dy\int^{y}dx\,
\frac{a_{1}+\frac{a_{0}}{x}+\frac{a_{-1}}{x^{2}}}
{\sqrt{(x_{1}-x)(x-x_{2})(x-x_{3})(x-x_{4})}}\nonumber\\\nonumber\\
&=&\frac{(1-2\alpha)(x_{1}-x_{4})}
{\pi\sqrt{(x_{1}-x_{3})(x_{2}-x_{4})}}
{\mathrm{\Pi}}\left(\omega^{2},r\right)
-\frac{x_{2}x_{3}(x_{1}-x_{4})}
{\pi\sqrt{(x_{1}-x_{3})(x_{2}-x_{4})x_{1}x_{2}x_{3}x_{4}}}
{\mathrm{\Pi}}\left(\omega_{1}^{2},r\right)
\nonumber\\\nonumber\\\label{eq:splus} 
&-&\frac{x_{2}(x_{1}-x_{3})(x_{1}-x_{4})
{\mathrm{K}}(r)}
{2\pi\sqrt{(x_{1}-x_{3})(x_{2}-x_{4})x_{1}x_{2}x_{3}x_{4}}}
\left(1-\frac{x_{1}(x_{2}-x_{4})}{x_{2}(x_{1}-x_{4})}
\frac{{\mathrm{E}}(r)}{{\mathrm{K}}(r)}\right)+\alpha\,,
\eea
and
\bea
{\mathcal{S}}_{-}&=&\frac{1}{\pi}\int_{x_{4}}^{x_{3}}dy\int_{y}dx\,
\frac{a_{1}+\frac{a_{0}}{x}+\frac{a_{-1}}{x^{2}}}
{\sqrt{(x_{1}-x)(x_{2}-x)(x_{3}-x)(x-x_{4})}}\nonumber\\\nonumber\\
&=&\frac{(1-2\alpha)(x_{1}-x_{4})}{\pi\sqrt{(x_{1}-x_{3})(x_{2}-x_{4})}}
{\mathrm{\Pi}}\left(\tilde{\omega}^{2},r\right)
-\frac{x_{2}x_{3}(x_{1}-x_{4})}
{\pi\sqrt{(x_{1}-x_{3})(x_{2}-x_{4})x_{1}x_{2}x_{3}x_{4}}}
{\mathrm{\Pi}}\left(\tilde{\omega}_{1}^{2},r\right)
\nonumber\\\nonumber\\\label{eq:sminus}  
&+&\frac{x_{3}(x_{1}-x_{4})(x_{2}-x_{4}){\mathrm{K}}(r)}
{2\pi\sqrt{(x_{1}-x_{3})(x_{2}-x_{4})x_{1}x_{2}x_{3}x_{4}}}
\left(1-\frac{x_{4}(x_{1}-x_{3})}{x_{3}(x_{1}-x_{4})}
\frac{{\mathrm{E}}(r)}{{\mathrm{K}}(r)}\right)+\alpha\,.
\eea 
Here the elliptic moduli $\omega^{2}$, $\omega_{1}^{2}$,
$\tilde{\omega}^{2}$ and $\tilde{\omega}_{1}^{2}$ are given by 
\be
\omega^{2}=-\frac{(x_{3}-x_{4})}{(x_{1}-x_{3})}\,,\;\;\;\;\;
\omega_{1}^{2}=-\frac{x_{1}(x_{3}-x_{4})}{x_{4}(x_{1}-x_{3})}\,,
\;\;\;\;\; \tilde{\omega}^{2}=\frac{r}{\omega^{2}}\,,\;\;\;\;\;
\tilde{\omega}_{1}^{2}=\frac{r}{\omega_{1}^{2}}\,. 
\ee 
In deriving ${\mathcal{S}}_{+}$ and ${\mathcal{S}_{-}}$, we have
repeatedly used the identities (A.3) and (A.4) in the Appendix A to 
simplify the expressions. The appearance of elliptic integrals of the
third kind in these expressions for the filling ratio is rather
suggestive since we have also encountered these in the semiclassical
string computation.
From the identity (A.5), we can deduce that
there are in fact only three independent elliptic moduli in the
expressions namely, $r$, $\omega^{2}$ and $\omega_{1}^{2}$. This is
exactly the same number as in the semiclassical string analysis. 
It is also easily checked that 
that ${\mathcal{S}}_{+}+{\mathcal{S}}_{-}=\alpha$. 

While the condition
${\mathcal{S}}_{+}={\mathcal{S}}_{-}$ can be trivially satisfied in
the situation with symmetrical cuts, in the deformed case equating the
expressions (\ref{eq:splus})
and (\ref{eq:sminus}) yields a
non-trivial constraint on the end points of the cuts. 
Although the two cuts still contain equal numbers of Bethe roots, the
potential felt along each cut is now different and the roots can have
different distributions. The two branch cuts are therefore  
not necessarily symmetrical or of equal length.

\subsection{Matching the string and spin-chain parameters}
At this point we can make further concrete 
connections with the semiclassical string analysis and relate
the elliptic parameters in the two pictures.
In this context the key equation is the vanishing A-cycle condition
(\ref{eq:acycle}) which 
can be rewritten using the identity (A.3) as an expression for the
filling ratio
\be
\alpha=\frac{1}{2}-\frac{(x_{1}x_{2}+x_{3}x_{4})}{4\sqrt{x_{1}x_{2}x_{3}x_{4}}}
+\frac{(x_{1}-x_{4})(x_{2}-x_{3})}{4\sqrt{x_{1}x_{2}x_{3}x_{4}}}
\frac{{\mathrm{E}}\left[-\frac{r}{1-r}\right]}
     {{\mathrm{K}}\left[-\frac{r}{1-r}\right]}\,. 
\ee 
Comparing this with the corresponding string theory expression
(\ref{eq:filling}) we see that they are rather similar. Indeed, for
the correspondence between the spin-chain and semiclassical string
solutions to work, they have to be equal! 
This requires us to naturally identify the moduli from
the two different calculations as follows: 
\bea
x_{+}^{(0)}&=&\frac{1}{2}-\frac{(x_{1}x_{2}+x_{3}x_{4})}
{4\sqrt{x_{1}x_{2}x_{3}x_{4}}}\,,\;\;\;\;\;
x_{-}^{(0)}=\frac{1}{2}-\frac{(x_{1}x_{3}+x_{2}x_{4})}
{4\sqrt{x_{1}x_{2}x_{3}x_{4}}}
\,,\nonumber\\
x_{0}^{(0)}&=&\frac{1}{2}-\frac{(x_{1}x_{4}+x_{2}x_{3})}
{4\sqrt{x_{1}x_{2}x_{3}x_{4}}}\,,\;\;\;\;\;
k^{(0)}=-\frac{(x_{1}-x_{2})(x_{3}-x_{4})}
{(x_{1}-x_{4})(x_{2}-x_{3})}\,.
\eea
As an elementary check for these expressions, we notice that in the
limit of two symmetrical cuts, $x_{4}\to -x_{1}$ and $x_{3}\to
-x_{2}$, which was the natural Bethe roots distribution to consider
when $\beta$ the deformation parameter is set to zero,
\be
x_{+}^{(0)}= 1\,,\;\;\;\;\;
x_{-}^{(0)}= 0\,,\;\;\;\;\;
x_{0}^{(0)}=k^{(0)}_{0}= -\frac{(x_{1}-x_{2})^{2}}{4x_{1}x_{2}}\,.
\ee
Here we have followed the convention in \cite{Kazakov:2004qf}
so that $\sqrt{x_{1}x_{2}x_{3}x_{4}}\to -x_{1}x_{2}$ in the limit.
The first two expressions above simply give the correct constants for
the 
undeformed case, whereas the third expression coincides with the
identification between the elliptic moduli in the original folded
string and spin chain calculations given in $\cite{Arutyunov:2003uj}$.

Furthermore, for the semiclassical string solution in the previous
section to be valid, the Virasoro constraint (\ref{eq:vir}) has to be
satisfied to all orders in $\beta$. Using the identification of the moduli
above, we can re-express the lowest order Virasoro constraint
equation as \be
\left(\sum_{i=1}^{4}x_{i}\right)\left(\sum_{i=1}^{4}\frac{1}{x_{i}}\right)=0\,.
\ee 
This condition imposes an extra constraint on the end points
$x_{1}\dots x_{4}$ in order for the correspondence between the spin
chain and the semiclassical string to be consistent. 
In the case of the undeformed
symmetrical cuts distribution, the constraint is trivially satisfied.
However now we need a nontrivial relation between the branch point
locations which leads to either $\sum_{i=1}^{4}x_{i}=0$ or
$\sum_{i=1}^{4}\frac{1}{x_{i}}=0$ so that the Virasosro constraints
are satisfied. As we see below the latter condition $\sum_{i=1}^4
1/x_i=0$ is also the necessary condition for the  
matching of the anomalous dimension calculated from the two sides of
the correspondence.

We can now calculate the anomalous dimension $\gamma$ using the
definition (\ref{eq:gamma}) after using the identity (A.5) which yields
\be \gamma=\frac{1}{32\pi^{2}}
\left\{\left(\frac{1}{x_{2}}-\frac{1}{x_{4}}\right)
\left(\frac{1}{x_{1}}-\frac{1}{x_{3}}\right)
\frac{{\mathrm{E}}(r)}{{\mathrm{K}}(r)}
-\frac{\left(\left(\frac{1}{x_{1}}-\frac{1}{x_{3}}\right)+
\left(\frac{1}{x_{2}}-\frac{1}{x_{4}}\right)\right)^{2}}{4}\right\}\,.
\ee
This expression, albeit simple, at first sight looks nothing like the corresponding
expression for the energy $\varepsilon_{1}$ we derived earlier
(\ref{eq:dim}) from
the string solution. But using (\ref{eq:kr}) for ${\mathrm{K}}(r)$ and
the elliptic modular 
transformations (A.3) we can massage the expression above into the form
\bea
\gamma&=&\frac{2{\mathrm{K}}\left(\frac{-r}{1-r}\right)
\left\{{\mathrm{E}}\left(\frac{-r}{1-r}\right)-
\left(\frac{1}{r}\right){\mathrm{K}}
\left(\frac{-r}{1-r}\right)\right\}}{\pi^{2}}
-\frac{\left(\left(\frac{1}{x_{1}}+\frac{1}{x_{4}}\right)-
\left(\frac{1}{x_{2}}+
\frac{1}{x_{3}}\right)\right)^{2}}{128\pi^{2}}
\nonumber\\
&=&\frac{2{\mathrm{K}}\left(k^{(0)}\right)
\left\{{\mathrm{E}}\left(k^{(0)}\right)-\left(1-k^{(0)}\right)
{\mathrm{K}}\left(k^{(0)}\right)\right\}}{\pi^{2}}
-\frac{\left(\left(\frac{1}{x_{1}}+\frac{1}{x_{4}}\right)-
\left(\frac{1}{x_{2}}+\frac{1}{x_{3}}\right)
\right)^{2}}{128\pi^{2}}\nonumber\,.\\
\eea 
In the second line we have used the identification
of the moduli in (5.25). Now as we can see, $\gamma$ looks
remarkably similar to $\varepsilon_{1}$ (\ref{eq:dim}) and the two
would be identical if 
we could set the second term in $\varepsilon_{1}$ and in $\gamma$
equal to each other. Equating these two using (\ref{eq:kr}) and (5.25)
we find, 
\be
\frac{2{\mathrm{K}}\left(k^{(0)}\right)^{2}
\left(x_{+}^{(0)}+x_{-}^{(0)}-1\right)}
{\pi^{2}\left(x_{+}^{(0)}-x_{-}^{(0)}\right)}
=\frac{\left(\frac{1}{x_{1}}+\frac{1}{x_{3}}\right)
\left(\frac{1}{x_{2}}+\frac{1}{x_{4}}\right)}{32\pi^{2}}\,.
\ee 
The necessary condition for the semiclassical string and
spin-chain calculations to match then becomes 
\be
4\left(\frac{1}{x_{1}}+\frac{1}{x_{3}}\right)
\left(\frac{1}{x_{2}}+\frac{1}{x_{4}}\right)=
\left(\left(\frac{1}{x_{1}}+\frac{1}{x_{4}}\right)-
\left(\frac{1}{x_{2}}+\frac{1}{x_{3}}\right)\right)^{2}\,,
\ee 
But this is precisely the condition $\sum_{i=1}^{4}\frac{1}{x_{i}}=0$
which satifies the second Virasoro constraint! 
This striking match between the spinning string analysis
and the twisted spin chain calculation for the anomalous dimension is
the main result of our paper.

\section{Summary and future direction}

In this paper, we have explicitly constructed a two spin 
semiclassical folded string solution in the Lunin-Maldacena background. This
solution is a folded configuration with nontrivial winding. 
It should be regarded
as the natural deformation of the two-spin semiclassical folded
string solution in the original $AdS_{5}\times S^{5}$ under the effect
of non-trivial deformation parameter $\beta$. We then calculated its
energy to the lowest order in an expansion in powers of
$\frac{\lambda}{J^{2}}$. This then gives a 
non-trivial $\beta$ dependent correction to the energy of the original
semiclassical strings calculated in \cite{Frolov:2003xy}.

We then performed an explicit twisted spin chain analysis for a
general two-cut Bethe roots distribution. We naturally identified the
$\beta$ dependent end points of the cuts with the moduli in the
elliptic functions characterizing the folded string solution. With
these identifications, the anomalous dimension calculated using the
twisted spin chain was shown to precisely coincide with the energy of
the folded string, after we impose the appropriate Virasoro
constraint. The striking match provides another non-trivial check for
the (one loop) integrability in the less supersymmetric
backgrounds. Effectively, 
we have also demonstrated the role played by the second Virasoro
constraint in this analysis. (Some earlier discussions can be found in
\cite{Arutyunov:2003za}.).

A natural extension of this work would be trying to understand 
how conserved charges other than the energy can be matched
between the twisted spin chain and spinning strings, as was done 
in \cite{Arutyunov:2003rg}. It is known that the twisted
spin chain is an integrable system with a large number of conserved
charges. On the other hand the class of semiclassical string
solutions we have in this paper belongs to the so-called
Neumann-Rosochatius integrable system \cite{Arutyunov:2003za}. It
would be both physically and mathematically interesting to see how the
conserved charges in the twisted spin chain can also arise in this class
of semiclassical solutions. Following \cite{Arutyunov:2003rg}, this
would presumably require a generalization of the B\"acklund
transformation to the semiclassical string solution with non-trivial
winding number. We leave this line of investigation for future work. 
\newline

{\bf Acknowledgements}: We have benefitted from valuable discussions with
Nick Dorey, Nick Manton and Benoit Vicedo. 
We also appreciate the useful comments from Sergey Frolov, Radu Roiban and Arkady Tseytlin.
H.Y.C. would also like to thank National Taiwan University for the
hospitality during part of this work. The work of H.Y.C. is supported
by a Benefactors' Scholarship from St. John's College, Cambridge.
S.P.K. is supported by a PPARC Advanced Fellowship.
\newline

\startappendix
\Appendix{Definitions and identities for the complete
elliptic functions}

Here we list the convention of the complete elliptic integrals
used in this paper: \bea
&&{\mathrm{K}}[k]=\int^{\frac{\pi}{2}}_{0}\frac{d\phi}{\sqrt{1-k\sin^{2}\phi}}\,,
\;\;\;\;\;{\mathrm{E}}[k]=\int^{\frac{\pi}{2}}_{0}d\phi\,\sqrt{1-k\sin^{2}\phi}\,\\
&&{\mathrm{\Pi}}(\omega^{2},k)=\int^{\frac{\pi}{2}}_{0}\frac{d\phi}{(1-\omega^{2}\sin^{2}\phi)\sqrt{1-k\sin^{2}\phi}}\,.
\eea
In addition, some very useful identities for identifying the
moduli in the sigma model and spin chain calculations are \bea
&&{\mathrm{K}}[k]=\frac{1}{\sqrt{1-k}}{\mathrm{K}}\left[\frac{-k}{1-k}\right]\,,\;\;\;\;\;\;
{\mathrm{E}}[k]=\sqrt{1-k}\,{\mathrm{E}}\left[\frac{-k}{1-k}\right]\,,\\
&&{\mathrm{\Pi}}(\omega^{2},k)
=\frac{1}{(1-\omega^{2})\sqrt{1-k}}{{\mathrm{\Pi}}}\left(-\frac{\omega^{2}}{1-\omega^{2}},-\frac{k}{1-k}\right)\,,\\
&&{\mathrm{\Pi}}(\omega^{2},k)+{\mathrm{\Pi}}(k/\omega^{2},k)={\mathrm{K}}[k]
+\frac{\pi}{2}\sqrt{\frac{\omega^{2}}{(1-\omega^{2})(\omega^{2}-k)}}\,.
\eea The Legendre relation is given by \be
{\mathrm{E}}(k){\mathrm{K}}(1-k)+{\mathrm{E}}(1-r){\mathrm{K}}(r)-{\mathrm{K}}(r){\mathrm{K}}(1-r)=\frac{\pi}{2}\,.\ee

\Appendix{Explicit derivation of the folded string energy}

Here we describe how one can obtain the first order correction to
the folded string energy $\varepsilon_{1}$ in (4.30).

Starting from the Virasoro constraint (4.15), the energy $E$, the
total angular momentum $J$ and the t' Hooft coupling $\lambda$ are
related to $\kappa$ and ${\mathcal{J}}$ by $E=\sqrt{\lambda}\kappa$
and $J=\sqrt{\lambda}{\mathcal{J}}$. Using the expansions (4.29), we
can expand $\kappa$ to the first order $\frac{1}{{\mathcal{J}}}$ \be
\kappa={\mathcal{J}}+\frac{1}{{\mathcal{J}}}\left(\frac{\left(\alpha-x_{0}^{(0)}\right)}{(1-\alpha)}{\mathcal{F}}
+\frac{2{\mathrm{K}}\left(k^{(0)}\right)^{2}\left(x_{+}^{(0)}+x_{-}^{(0)}
-1\right)}{\pi^{2}\left(x_{+}^{(0)}-x_{-}^{(0)}\right)}\right)\,.
\ee Here we have also used the zeroth order relation (4.31) for
$\alpha$ to simplify the expression. The function ${\mathcal{F}}$ is
given by \bea {\mathcal{F}}&=&
\frac{x_{+}^{(2)}}{\alpha}\left(1-\frac{{\mathrm{E}}\left(k^{(0)}\right)}{{\mathrm{K}}\left(k^{(0)}\right)}\right)
+\frac{x_{-}^{(2)}}{\alpha}\frac{{\mathrm{E}}\left(k^{(0)}\right)}{{\mathrm{K}}\left(k^{(0)}\right)}\nonumber\\
&+&\frac{k^{(2)}\left(x_{+}^{(0)}-x_{-}^{(0)}\right)}{2\alpha\left(1-k^{(0)}\right)k^{(0)}}
\left(\left(1-k^{(0)}\right)\left(1-\frac{{\mathrm{E}}\left(k^{(0)}\right)}{{\mathrm{K}}\left(k^{(0)}\right)}\right)^{2}
+k^{(0)}\left(\frac{{\mathrm{E}}\left(k^{(0)}\right)}{{\mathrm{K}}\left(k^{(0)}\right)}\right)^{2}\right)\,.
\eea To work out $x_{+}^{(2)}$, $x_{-}^{(2)}$ and $k^{(2)}$ in terms
of $x_{+}^{(0)}$, $x_{-}^{(0)}$, and $k^{(0)}$, one can in principle
expand (4.25), (4.26) and the sum of (4.27) and (4.28) to the first
order in $\frac{1}{{\mathcal{J}}}$ and solve the simultaneous
equations. However, we are only interested in $\kappa$ here, let us
focus on the expansion of (4.25) at the order
$\frac{1}{{\mathcal{J}}}$, \bea
&&\frac{-4{\mathrm{K}}\left(k^{(0)}\right)^{2}}{\pi^{2}\left(x_{+}^{(0)}-x_{-}^{(0)}\right)}=
\frac{2x_{+}^{(2)}}{\alpha(1-\alpha)}\left(1-\frac{{\mathrm{E}}\left(k^{(0)}\right)}{{\mathrm{K}}\left(k^{(0)}\right)}\right)
+\frac{2x_{-}^{(2)}}{\alpha(1-\alpha)}\frac{{\mathrm{E}}\left(k^{(0)}\right)}{{\mathrm{K}}\left(k^{(0)}\right)}\nonumber\\
&&+\frac{k^{(2)}\left(x_{+}^{(0)}-x_{-}^{(0)}\right)}{\alpha(1-\alpha)\left(1-k^{(0)}\right)k^{(0)}}
\left(\left(1-k^{(0)}\right)\left(1-\frac{{\mathrm{E}}\left(k^{(0)}\right)}{{\mathrm{K}}\left(k^{(0)}\right)}\right)^{2}
+k^{(0)}\left(\frac{{\mathrm{E}}\left(k^{(0)}\right)}{{\mathrm{K}}\left(k^{(0)}\right)}\right)^{2}\right)\,.
\eea the expression on the right hand side is proportional to
${\mathcal{F}}$! We can substitute the left hand side into
${\mathcal{F}}$ and $\kappa$, hence the simple expression for
$\varepsilon_{1}$ in (4.30).

\end{document}